\documentclass[fleqn,a4paper]{vch-book}
\usepackage[authoryear]{natbib}
\usepackage{amsmath}
\usepackage{amssymb}
\usepackage{makeidx}
    \makeindex
\usepackage{times}
\usepackage{tabularx}
\usepackage{booktabs}
\usepackage[dvips]{epsfig}
\DeclareGraphicsExtensions{.eps,.ps,.pdf}

\chardef\bslash=`\\ 

\newcommand{\bibtex}{\ifx\is@itshape\f@shape{\fontshape{scit}\selectfont
Bib}\else\textsc{Bib}\fi\kern-.1em\TeX}

\title{Phase coherence length of an elongated Bose-Einstein
condensate}
\author{F. Gerbier \footnote{e-mail: fabrice.gerbier@iota.u-psud.fr}, S. Richard, J. H. Thywissen, M.
Hugbart, P. Bouyer, and A. Aspect} \affil{Groupe d'Optique
Atomique, Laboratoire Charles Fabry de l'Institut d'Optique
\footnote{UMRA 8501 du CNRS}, B\^{a}t. 503, Campus universitaire
d'Orsay, , F-91403 ORSAY, FRANCE}

\begin{document}
\maketitle

The experimental study of one dimensional (1D) degenerate Bose
gases, where the radial motion is``frozen'', is currently an
important direction of research in ultracold atom physics
\cite{gorlitz01,schreck01,greiner01}. In uniform 1D systems,
fluctuations of the condensate wavefunction around equilibrium are
pronounced, because of a large population of low-lying states; as
a result, no long range order exists at any temperature. In a
trapped 1D gas, the finite size of the sample naturally introduces
a low-momentum cutoff, and at sufficiently low temperature a phase
coherent sample can exist (see \cite{petrov1d} and references
therein for the properties of 1D gases). Essentially, the same
analysis holds for three dimensional (3D) trapped gases in very
elongated traps \cite{petrov3d}: here the interaction energy per
particle is larger than the radial frequency, and the condensate
wavefunction is built from several radial modes. Still, low-energy
excitations with a frequency smaller than the radial frequency are
effectively one-dimensional, and induce large fluctuations of the
condensate phase \cite{petrov3d}. Such condensates with a
fluctuating phase, or {\it{quasicondensates}}, were recently
observed experimentally in \cite{dettmer01,hellweg01}. In this
paper, we first discuss how to measure the velocity distribution
{\it{via}} Bragg transitions, and why this is an (indirect)
measurement of the coherence length. Then we present our
measurements \cite{richard02} for very elongated traps, that agree
well with the theory developed by Petrov {\em et al.}
\cite{petrov3d,gerbier02}.

Phase coherence of the (quasi)condensate
can be characterized by the $1/e$ decay
length of the spatial correlation function, $\mathcal{C}^{(1)}(s)=1/N
\int d^{3}{\mathbf{R}} \langle \Psi^{\dagger}({\mathbf{R}}+s
{\mathbf{e_{\mathrm{z}}}}/2) \Psi( {\mathbf{R}}-s
\mathbf{e_\mathrm{z}}/2) \rangle$ \cite{zambelli00}, with $\Psi$ the
atomic field operator.
Consider a sample of $N_{0}$ condensed atoms, trapped in a
cylindrically symmetric, harmonic trap with an aspect ratio
$\lambda=\omega_{\mathrm{z}}/\omega_{\perp} \ll 1$, in the 3D
Thomas-Fermi regime.
Petrov {\em et al.} \cite{petrov1d,petrov3d} investigated the long wavelength
fluctuations of the phase of $\Psi$, with the following conclusions.
Below a characteristic temperature $T_{\phi}=15 N_{0} (\hbar
\omega_{\mathrm{z}})^{2}/32\mu$, the phase fluctuations are small,
and the coherence length is essentially the condensate size $L$,
as in ``ordinary'' 3D condensates \cite{stenger99,hagley99} (to
put it differently, $\mathcal{C}^{(1)}$ is limited by the density
envelope of $\Psi$). On the other hand, for $T \gg T_{\phi}$ the
coherence length is limited to a value much smaller than $L$, the
value at the center of the trap being $L_{\phi}= L T_{\phi}/T$.

\begin{figure}[c]
\includegraphics[width=12cm,height=8cm]{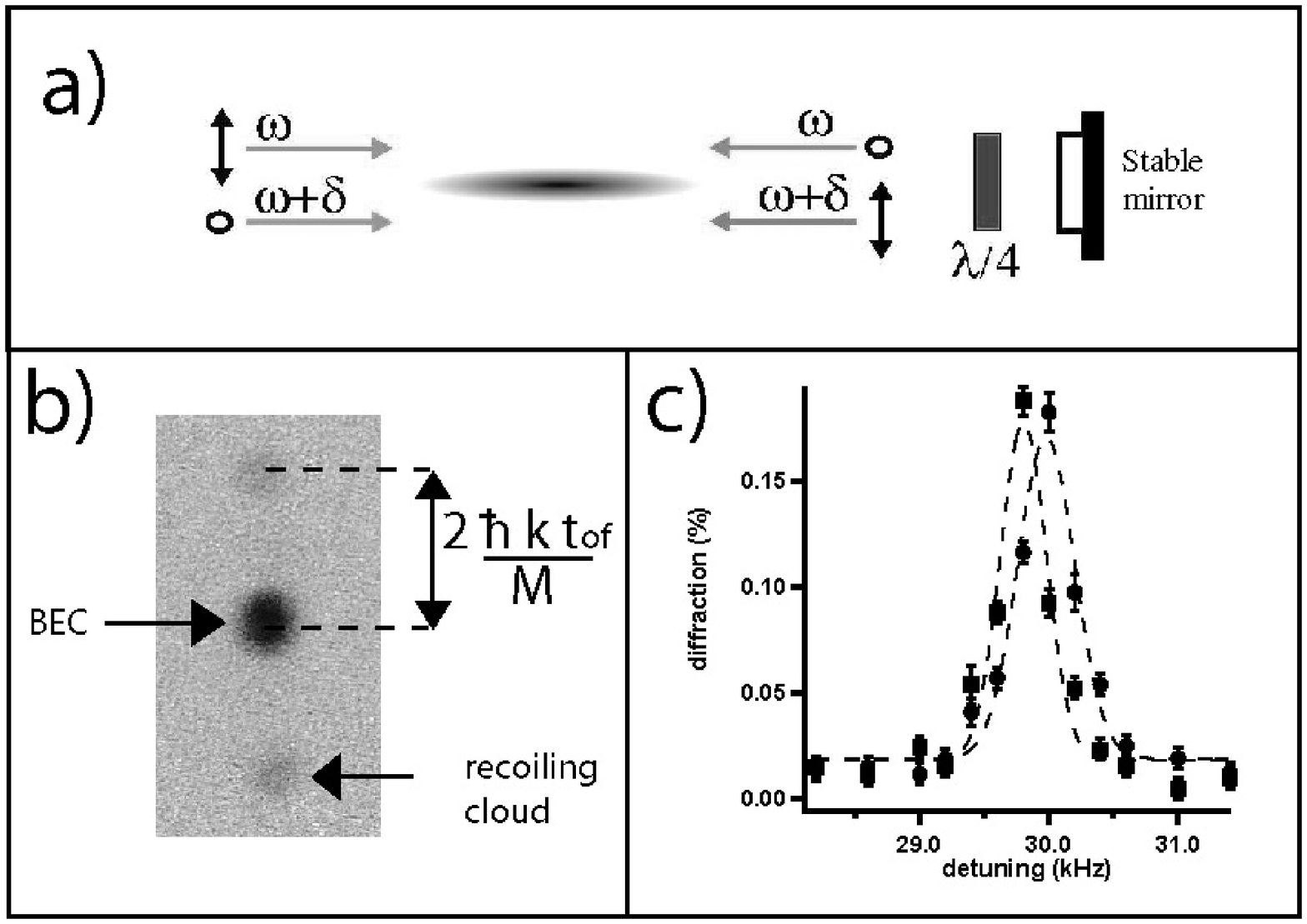}
\caption{a) Schematics of the optical setup for Bragg
spectroscopy, as done in our experiment (see text for details). b)
Density distribution after a 20\,ms time of flight. The diffracted atoms
are clearly seen as the two peaks recoiling from the condensate.
c) A typical spectrum, recording the number of recoiling atoms in
each peak as a function of relative detuning between the Bragg
beams. } \label{fig01}
\end{figure}

A convenient technique to measure the coherence length is Bragg
Spectroscopy, as was done in \cite{stenger99} in the radial
direction. One illuminates the atomic sample with
counter-propagating laser beams with frequency difference $\omega$
and wavevectors $\pm {\mathbf{k_{\mathrm{L}}}}$, parallel to the
long axis $z$ of a cigar-shaped condensate (see Fig.~\ref{fig01}).
The basic process involves the absorption of a photon in one beam
and stimulated emission into the other, with a net momentum and
energy transfer $\pm 2 \hbar {\mathbf{k_{\mathrm{L}}}}$ and $\hbar
\omega$, respectively. This process is resonant around the
frequency that matches the gain in kinetic energy: for an atom at
rest it is $\omega=4\hbar k_{\rm{L}}^{2}/2M=\omega_{0}$. If the
initial velocity is non-zero, then the resonance frequency is
Doppler-shifted according to $\omega=\omega_{0}+2
k_{\mathrm{L}}v_{\mathrm{z}}$.  Therefore, by scanning the
relative frequency of the two beams one scans the velocity
distribution $\mathcal{P}(v_{\mathrm{z}})$. Since $\mathcal{P}$ is
the Fourier transform of $\mathcal{C}^{(1)}(s)$ \cite{zambelli00},
the root mean square (rms) width of the spectrum (``Doppler
width'') can be used to infer the coherence length according to
$\Delta \omega^{\mathrm{D}}=2 k_{\mathrm{L}} v_{\mathrm{rms}} \sim
2 (\hbar k_{\mathrm{L}}/M L_{\phi})$, to within a numerical factor
that depends on the lineshape. In other words, we probe the phase
fluctuations of the wavefunction by measuring the momentum
distribution, {\em i.e.} by probing the spatial gradient of the
phase, rather than the phase itself .

Experimentally, axial Bragg spectroscopy directly in the trap is
difficult because of collisions between the diffracted atoms and
the parent BEC. First, mean-field interactions inhomogeneously
broaden the line (typically by $\sim 500$\,Hz in our case)
\cite{stenger99,zambelli00}. Second, the mean free path of the
diffracted atoms $l_{\mathrm{mfp}}$ is typically much smaller than
L, thus producing a diffuse background, rather than a well-defined
diffracted cloud \cite{chikkatur00}. We solve these problems by
opening the trap abruptly, and letting the BEC expand for 2\,ms
before applying the Bragg pulse: the density drops by a factor
$\sim 90$, and the mean field broadening and destructive
collisions are suppressed. The released mean field energy converts
mainly into radial velocity in a time $\omega_{\perp}$, while a
small axial velocity field develops \cite{castin96,kagan96}. For
our experimental parameters the latter is negligible ($\sim
40$\,Hz to be compared to the typical linewidth $\sim 200$\,Hz,
see below). The fluctuating velocity distribution is essentially
frozen during the release of the condensate, and the velocity
width remains $\sim \hbar/ML_{\phi}$ during the expansion
\cite{gerbier02}. For quantitative comparison with our data, we
have assumed that all contributions to the width add in
quadrature. A gaussian fit to the ``ideal'' velocity distribution
calculated in \cite{gerbier02} gives an effective rms width
$v_{\mathrm{\phi,rms}}\approx 0.62 \hbar/ML_{\phi}$.

\begin{figure}[t]
\includegraphics[height=4.5cm]{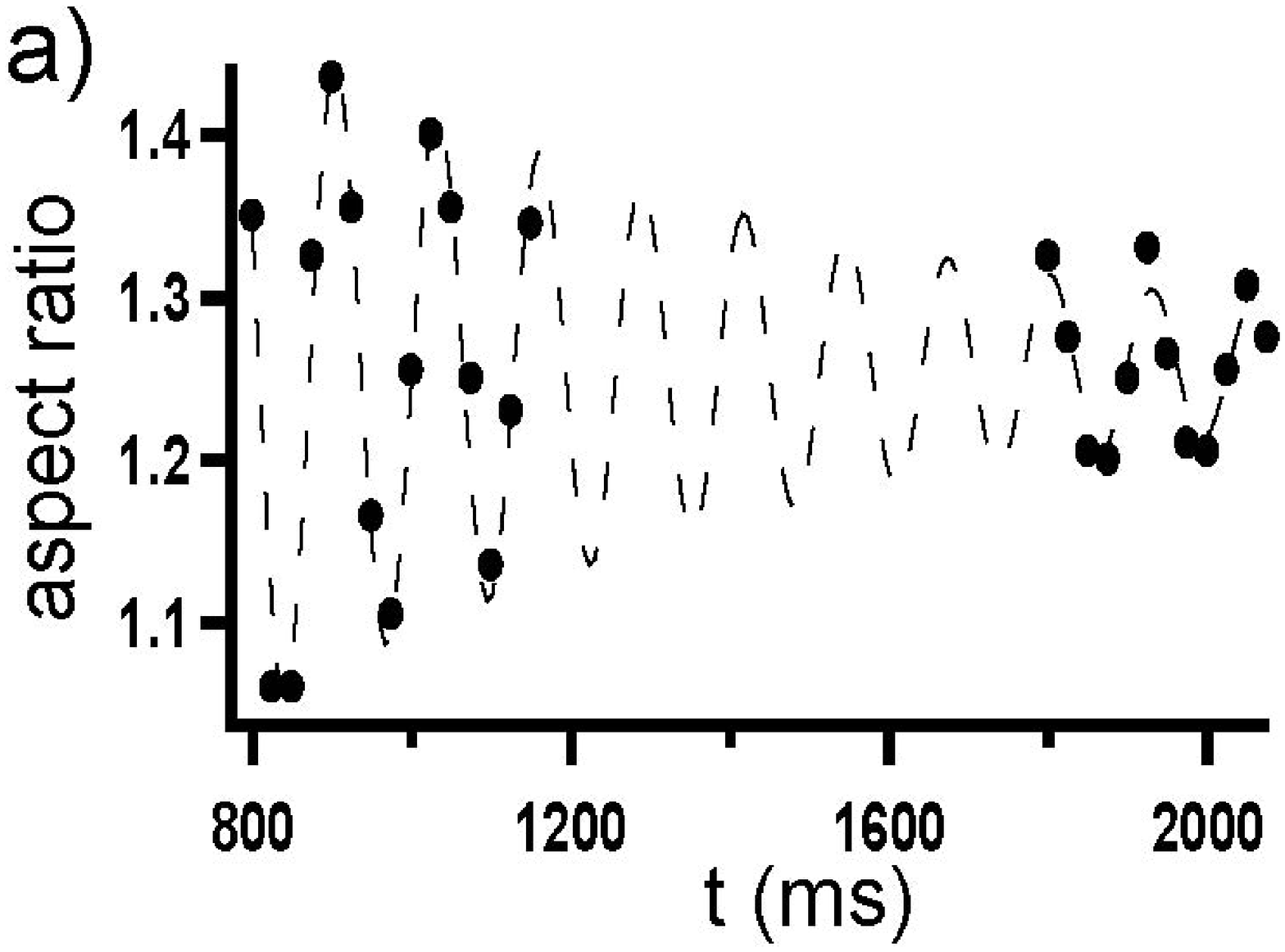}
\includegraphics[height=4.5cm]{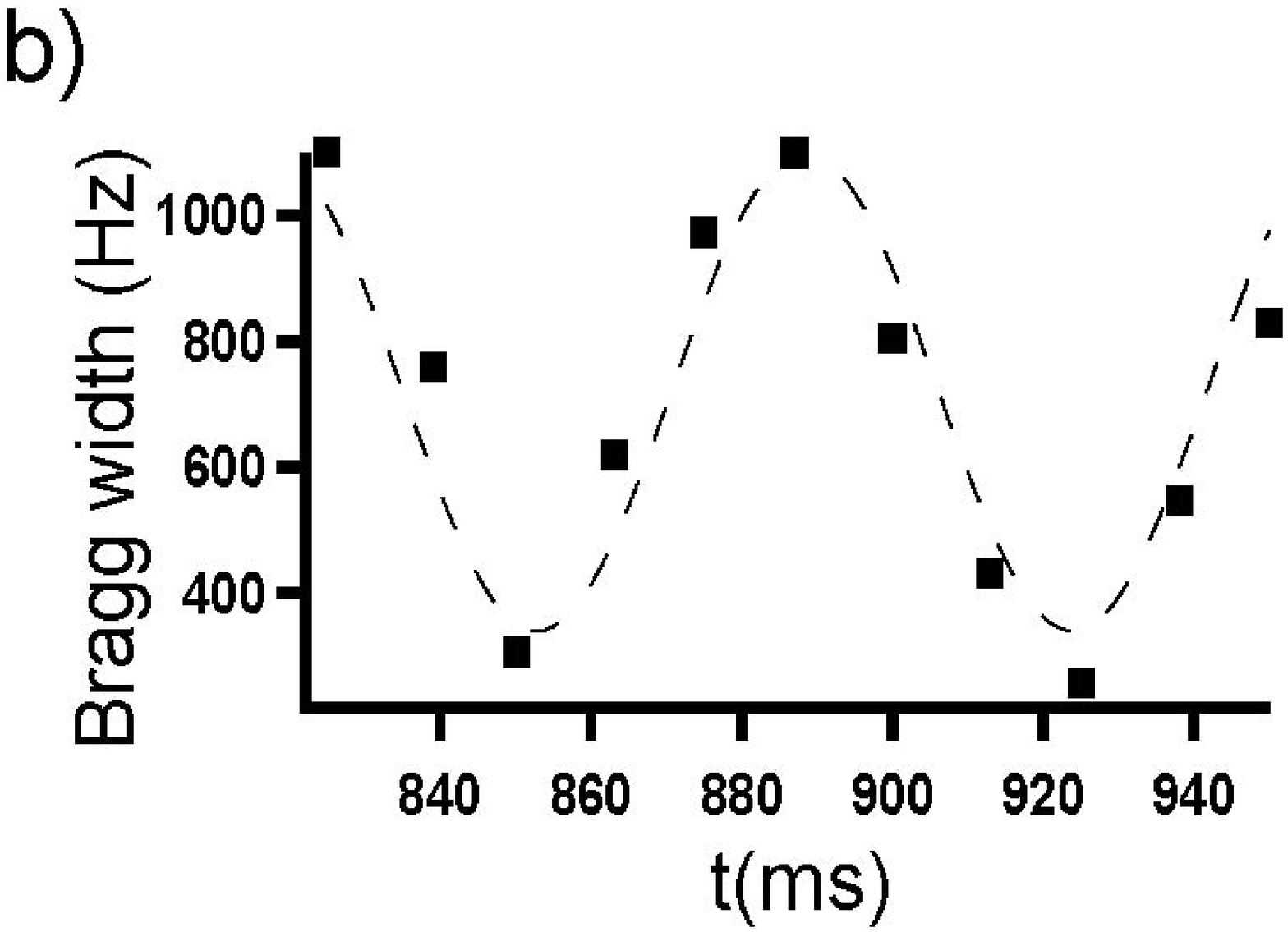}
\caption{Axial oscillations triggered by abrupt condensation (see
text). a) Oscillations in the aspect ratio of the condensate after a
24\,ms free expansion. The
dashed line is a damped sinusoid fit, yielding a frequency of
$7.8(3)$\,Hz, in good agreement with the expected value (1.58 times
the trap frequency $\nu_{z}=5.0$\,Hz \cite{dalfovo99}), and a $1/$e
decay time of $760$\,ms. b) The corresponding oscillation oscillation
of the velocity width, observed via Bragg spectroscopy. The
frequency is about twice the bare frequency (more precisely we
find $14.1(6)$\,Hz), because the width is not sensitive to whether the
cloud is contracting or expanding.}
\label{fig02}
\end{figure}

In our experiment, we produce a condensate of $^{87}$Rb in the
$|{\rm{F}} = 1 ; {\rm{m_F}}=-1\rangle$ hyperfine level. The
magnetic trap is an iron-core electromagnet producing a
Ioffe-Pritchard type potential. However, an important difference
with respect to the trap used in \cite{desruelle99,lecoq01} is the
possibility to compensate the bias field down to a few Gauss.
Together with a strong radial gradient of $1400$\,G/cm, this
allows to reach radial frequencies of $\sim 800$\,Hz. At the same
time the axial confinement is shallow, with a typical frequency
$\sim 5$\,Hz. The overall duty cycle, including laser-cooling and
evaporation phases, is typically $30$\,s. After the last
evaporation ramp, we observe the formation of a condensate that
exhibits strong shape oscillations (see \cite{dalfovo99} and
references therein), as shown in Fig.\ref{fig02}a. This is in line
with a recent observation by Schvarchuck {\it{et al.}}
\cite{schvarchuck02}, who propose the picture of a BEC locally in
equilibrium at each point on the axis, but globally out of
equilibrium. Indeed, the radial degrees of freedom are rapidly
thermalized, but axial motion damps on a much longer time scale.
We plot in Fig.~\ref{fig02}b the width of the Bragg resonance for
such a non-equilibrium condensate as a function of time. It also
reveals an oscillation, however at about twice the mode frequency,
because the velocity width is not sensitive to whether the
condensate contracts or expands. The amplitude in velocity space
is quite large and disturbs significantly momentum spectroscopy:
therefore we hold the condensate for a variable time (typically 7
seconds) in presence of an rf shield, so that the collective
oscillation damps completely.

Let us discuss briefly a few important experimental points. As
shown in Fig.~\ref{fig01}a, the beams that form the lattice travel
parallel to the long axis of the cloud. The beams are
retroreflected to form two standing waves with orthogonal
polarizations and opposite chirp directions: therefore two
distinct clouds with opposite velocities are seen after time of
flight (Fig.~\ref{fig01}b). Stabilizing the last mirror against
mechanical vibrations was crucial for achieving sufficient
resolution. We mounted the mirror on a massive ($30 \times 30
\times 2$\,cm) aluminium plate, oriented such that vibrations
essentially occur perpendicular to the optical path. The Bragg
pulse is applied for a duration of $2$\,ms at around $30.1$\,kHz
to stimulate four-photon scattering. We used second-order instead
of first-order (two-photon) Bragg scattering to increase the
seperation of the diffracted clouds from the parent condensate.
After a 20\,ms free flight, we measure the relative number of
diffracted atoms by absorption imaging. The temperature and
condensate energies are also measured from this image, after
correcting for a small probe angle ($\sim20$\,mrad) with respect
to the vertical. The temperature is carefully controlled during
each spectrum by controlling the trap depth to $2$\,kHz precision
(corresponding to $\sim 10$\,nK).

\begin{figure}[t]
\includegraphics[height=4.5cm]{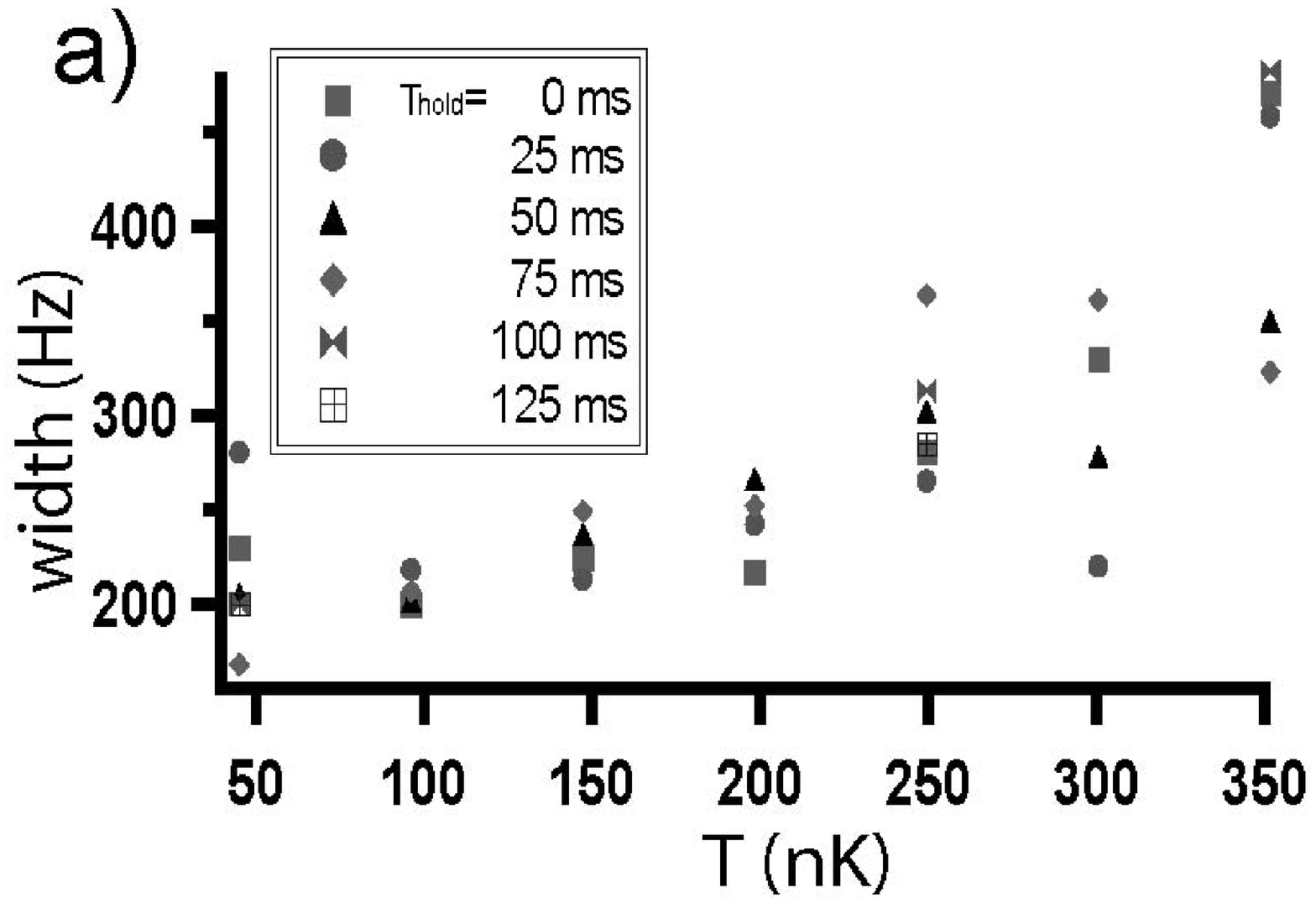}
\includegraphics[height=4.5cm]{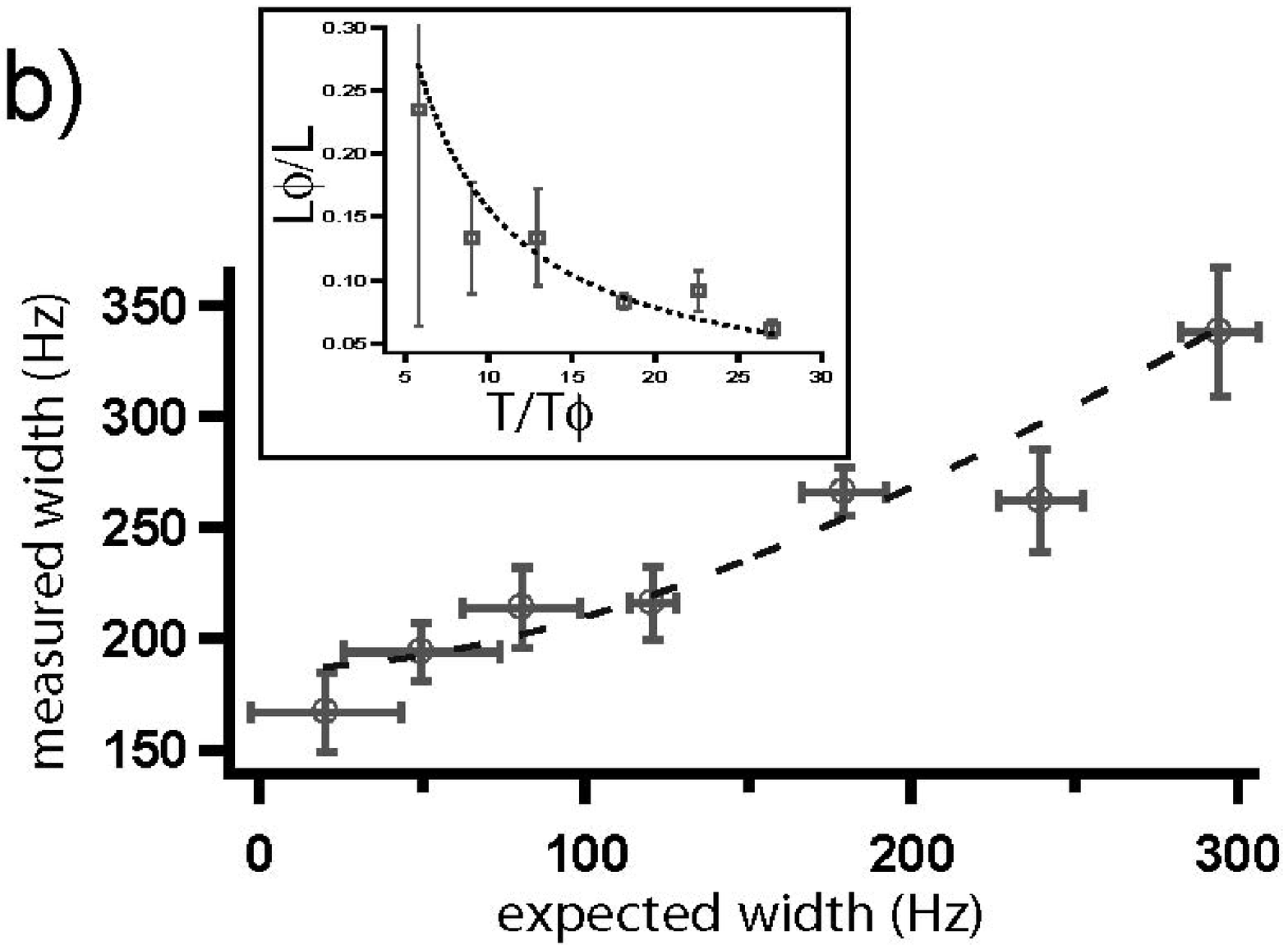}
\caption{a) Width of the Bragg resonance as a function of
temperature. The trapping geometry is $\omega_{\perp}=2 \pi \times
780$\,Hz, $\omega_{\mathrm{z}}=2 \pi \times 5.0$\,Hz and typically
$N_{0}= 5 \times 10^{4}$. For a given temperature, each point
correspond to a given delay before opening of the trap. The large
dispersion is due to a spurious center of mass motion and can be
corrected for (see text). b) Doppler width for elongated
condensates, corrected for the spurious sloshing, compared to the
theoretical expectation $\Delta_{\phi}=2 k_{\rm{L}} \times 0.62
\hbar/M L_{\phi}$. Dashed line is a fit with the form $(
\Delta_{0}^{2}+ \alpha^{2}\Delta_{\phi}^{2}) ^{1/2}$, yielding $
\Delta_{0}=185(9) $\,Hz (experimental resolution) and
$\alpha=0.96(8)$. In inset, we show the coherence length in an
elongated trap deduced from this measurement.} \label{fig03}
\end{figure}

We take Bragg spectra (diffraction efficiency versus detuning),
fixing the final rf frequency and therefore the number of
condensed atoms and the temperature of the surrounding thermal
cloud (see Fig.~\ref{fig01}c for a typical spectrum). Typically,
each point of the spectrum is averaged over five individual
measurements. Figure~\ref{fig03}a shows the Doppler width as a
function of temperature. A clear increasing trend is seen,
suggesting the presence of phase fluctuations in our sample. At
each temperature, between four and six spectra were taken, at a
variety of delays to sample any residual oscillations of the
trapped cloud.  A closer look at the resonance width as a function
of this delay (for a fixed temperature) reveals an oscillation at
the trap frequency. This is consistent with a center of mass
oscillation having a fluctuating {\it{amplitude}}, causing line
shifts random from shot to shot: averaging over many points
therefore results in line broadening. However, as the {\it{phase}}
of this oscillation is well defined, we are able to quantify the
broadening and correct for it. The Doppler width extracted after
correction of this spurious broadening is plotted in
Fig.~\ref{fig03}b versus the expected width, $\Delta_{\phi}=2
k_{\mathrm{L}} \times 0.62 \hbar /M L_{\phi}$ \cite{gerbier02}.
Using a quadrature fit, $\Delta^{2}= \Delta_{0}^{2}+ \alpha^{2}
\Delta_{\phi}^{2}$, with $\Delta_{0}$ the experimental resolution,
we find a slope $\alpha=0.96(8)$, sufficiently close to $1$ to
indicate good agreement with theoretical expectations. Thus, our
experiment provides a quantitative (and successful) test of the
theory of phase fluctuations in BEC developed in \cite{petrov3d}.
The coherence length deduced after substraction of the
experimental resolution is shown in inset of Fig.~\ref{fig03}b,
and seen to be substantially smaller than the condensate length.

Although quasicondensates have reduced long-range order, local
correlation properties are expected to be identical to those of a pure
condensate \cite{petrov1d,petrov3d}.
We test this prediction quantitatively by measuring the release
energy of the cloud. As discussed in Ref.~\cite{ketterle97}, if we
assume that the density fluctuations are $\delta n = C n_{0}$, we
find for the release energy $E_{\mathrm{rel}}=(1+C^2) 5\mu/7$,
with $\mu$ the chemical potential calculated for zero density
fluctuations and Thomas-Fermi condensate \cite{dalfovo99}. This
quantity depends (not sensitively) on atom number calibration,
which can be done by measuring the transition temperature: since
the explicit expression \cite{dalfovo99} depends only on the
properties of the thermal cloud above the transition, this is an
independent calibration. We measure the release energy by a
standard time of flight technique \cite{ensher96}, and from this
deduce $-0.03 \leq C \leq 0.05$, consistent with zero density
fluctuations, as for a true ({\it{\i.e.}} phase-coherent)
condensate. \footnote{Related work has been carried out at the
Institute for Quantum Optics in Hannnover (J.\ J.\ Arlt,
presentation at the meeting ``Recent developments in the physics
of cold atomic gases,'' Trento (2002)).}

In conclusion, we have measured the temperature-dependent
coherence length of an elongated Bose condensate. We find
quantitative agreement between our measurements and the
predictions of Shlyapnikov and coworkers
\cite{petrov1d,petrov3d,gerbier02}. In addition, our measurements
are consistent with the absence of density fluctuations, even in
the phase-fluctuating regime. Therefore they confirm the picture
of this new regime of quantum degeneracy in elongated traps: a
macroscopic wavefunction is still the correct concept, but its
phase is random, such that phase coherence does not extend across
the whole sample.

We gratefully acknowledge D. S. Petrov and G. V. Shlyapnikov for
many stimulating discussions. This work was supported by CNRS,
DGA, and EU. FG stay in Heidelberg was funded by the European
Science Fundation under the project BEC 2002+.


\end{document}